\documentclass[9pt,twocolumn,twoside]{osajnl}

\journal{ao} 

\usepackage{tikz,xcolor,hyperref}

\definecolor{lime}{HTML}{A6CE39}
\DeclareRobustCommand{\orcidicon}{
	\begin{tikzpicture}
	\draw[lime, fill=lime] (0,0) 
	circle [radius=0.16] 
	node[white] {{\fontfamily{qag}\selectfont \tiny ID}};
	\draw[white, fill=white] (-0.0625,0.095) 
	circle [radius=0.007];
	\end{tikzpicture}
	\hspace{-2mm}
}

\foreach \x in {A, ..., Z}{\expandafter\xdef\csname orcid\x\endcsname{\noexpand\href{https://orcid.org/\csname orcidauthor\x\endcsname}
			{\noexpand\orcidicon}}
}

\setboolean{shortarticle}{false}

\title{Simons Observatory: Broadband Metamaterial Anti-Reflection Cuttings for Large Aperture Alumina Optics}

\author[1,2*]{Joseph E. Golec\orcidA{}}
\author[1]{Shreya Sutariya\orcidG{}}
\author[2,3]{Rebecca Jackson}
\author[2]{Jerry Zimmerman}

\author[4]{Simon R. Dicker\orcidB{}}

\author[4]{Jeffrey Iuliano\orcidE}

\author[1,2,3,5,6]{Jeff McMahon}

\author[7,8]{Giuseppe Puglisi\orcidF{}}

\author[9]{Carole Tucker\orcidD{}}

\author[10]{Edward J. Wollack\orcidC{}}

\affil[1]{Department of Physics, University of Chicago, 5720 S Ellis Ave, Chicago, IL 60637, USA}
\affil[2]{Fermi National Accelerator Laboratory, Batavia, IL, 60510, USA}
\affil[3]{Department of Astronomy and Astrophysics, University of Chicago, 5640 S Ellis Ave, Chicago, IL 60637, USA}
\affil[4]{Department of Physics and Astronomy, University of Pennsylvania, 209 South 33rd Street, Philadelphia, PA, 19104, USA}

\affil[5]{Kavli Institute for Cosmological Physics, University of Chicago, 5640 S Ellis Ave, Chicago, IL 60637, USA}
\affil[6]{Enrico Fermi Institute, University of Chicago, 5640 S Ellis Ave, Chicago, IL
60637, USA}

\affil[7]{Dipartimento di Fisica, Universit\`a di Roma Tor Vergata, Via della Ricerca Scientifica, 1, 00133, Roma, Italy}

\affil[8]{INFN Sezione di Roma, Universit\`a di Roma Tor Vergata, Via della Ricerca Scientifica, 1, 00133 Roma, Italy}

\affil[9]{School of Physics and Astronomy, Cardiff University, The Parade, Cardiff, CF24 3AA, UK}

\affil[10] {NASA/Goddard Space Flight Center, Greenbelt, MD, USA 20771}

\affil[*]{Corresponding author: golecjoe@uchicago.edu}

\begin{abstract}
We present the design, fabrication, and measured performance of metamaterial Anti-Reflection Cuttings (ARCs) for large-format alumina filters operating over more than an octave of bandwidth to be deployed on the Simons Observatory (SO). The ARC consists of sub-wavelength features diced into the optic's surface using a custom dicing saw with near-micron accuracy. The designs achieve percent-level control over reflections at angles of incidence up to 20$^\circ$. The ARCs were demonstrated on four 42 cm diameter filters covering the 75-170 GHz band and a 50 mm diameter prototype covering the 200-300 GHz band. The reflection and transmission of these samples were measured using a broadband coherent source that covers frequencies from 20 GHz to 1.2 THz. These measurements demonstrate percent-level control over reflectance across the targeted pass-bands and a rapid reduction in transmission as the wavelength approaches the length scale of the metamaterial structure where scattering dominates the optical response. The latter behavior enables the use of the metamaterial ARC as a scattering filter in this limit.
\end{abstract}

\setboolean{displaycopyright}{true}

\begin{document}

\maketitle

\section{Introduction}
\label{sec:Introduction}
Astronomical observations at millimeter and sub-millimeter wavelengths are key to understanding aspects of the universe ranging from probing the conditions in the first instant to understanding the star formation history and details of stellar births in our galaxy. Large format detector arrays have revolutionized observations at these wavelengths with state-of-the-art focal planes containing tens of thousands of broadband detectors. These focal planes have driven the need for large aperture optical elements that can operate at cryogenic temperatures.


Alumina is a ceramic material that can be machined into large diameter optics for millimeter wave bands and its use as an IR rejecting filter is critical to the function of next-generation instruments. For the optics considered in this publication, we sourced our low-dielectric loss type alumina from NTK Ceratek in Japan \cite{Ceratek_JP}. High-purity (>99.\%) sintered alumina exhibits low dielectric absorption at millimeter wavelengths, but has relatively high dielectric losses in the sub-millimeter \cite{Alford_LowLoss,Breeze_LowLoss,Afsar_LowLoss}. This property, combined with strong restrahlen bands at infrared wavelengths, makes alumina an ideal filter material by effectively rejecting infrared radiation by scattering it out of the optical path or absorbing and efficiently thermally conducting it to a cryogenic system without significant heating. However, alumina's high index of refraction ($n=\sqrt{\epsilon_r}=3.14$) is approximately constant in the microwave \cite{Inoue_thermalandopticalproperties,Yukithesis, Lamb_OpticalProperties} and causes 26\% of incident light to be reflected per surface in the absence of optical coatings. These reflections not only reduce the amount of transmitted power that reach the detectors, but can also cause unwanted optical effects such as degraded image quality and reduced polarization purity. Therefore, reliable high quality anti-reflection (AR) coatings are needed.

Coatings can be realized by adding materials with carefully selected dielectric properties and thicknesses to the surface of interest. In the simplest realization, a quarter-wavelength thick layer of material with dielectric index $\sqrt{n}$ can be added to the surface of a material with index $n$. This quarter wavelength coating can perfectly cancel reflections at a particular wavelength and provide acceptable performance over a moderate bandwidth. Broader band performance is possible with multiple layer coatings or by extending these to the continuum limit to realize a gradient index coating \cite{AR_Coating_Review}. In this work we focus on two-layer metamaterial cuttings that function as an homogeneous two-layer coating which can achieve percent-level control of reflectance over an octave of bandwidth when applied to alumina. 

Enormous efforts have gone into developing multi-layer AR coatings for alumina. This includes laminated plastics \cite{SPT-3G_AR}, laminated epoxy \cite{EpoxyAR,Bicep_Plastic_Lam_AR,Nitta_EpoxyARandLaserAR}, laminates of commercial materials \cite{Mullite_AR}, and laser machined metamaterial cuttings \cite{Laser_AR,Nitta_EpoxyARandLaserAR,Matsumura_LaserAR}. The laminates require identifying materials with the required indices of refraction, controlling changes in the thickness and index throughout the lamination process, and control of cryomechanical delamination. The latter poses a critical risk for applications which require a high reliability. The recent results on laser machining shows promise as they have demonstrated coatings on deployed optics though it has yet to control reflections at the percent-level across an octave of bandwidth.

We present a new approach whereby we use dicing saw blades to cut metamaterial structures, comprised of sub-wavelength stepped pyramids, into the alumina surface. This approach is based on our group's previous work on silicon ARCs \cite{Datta_Si_AR,Coughlin_si_AR,Golec_si_AR} which are now deployed on AdvACTPol \cite{Thornton_ACT_Overview}, CLASS \cite{CLASS_Design_Overview}, TolTEC \cite{TolTEC_Overview}, and the Simons Observatory \cite{SO_Instrument_Overview}, but it is necessary to expand this technology to alumina substrates as well to take advantage of the IR rejecting material properties inherent in alumina that are not present in silicon. We demonstrate this technique on the alumina filters for the Simons Observatory (SO), a CMB observatory that is currently being deployed in the Atacama Desert in Chile.

The organization of this paper is as follows: Section 2 contains a brief description of the designs of the ARCs for two dichroic observing bands centered at 90/150 GHz (the Mid-Frequency bands or MF) and 220/280 GHz (the Ultra High-Frequency bands or UHF). The fabrication procedure with the dicing saw is described in section 3 and meteorology of the machined coatings follow in section 4. We present measurements of the reflection and transmission of the MF and UHF ARCs in section 5, along with the description of the apparatus used to make those measurements. We conclude with a discussion of the performance of the ARCs, and the production of these optical elements for SO and future experiments \cite{CMBS4_sciencebook,CMBS4_techbook}.

\section{Design}
\label{sec:Design}

\begin{figure}[t!]
\centering
\includegraphics[width=\linewidth]{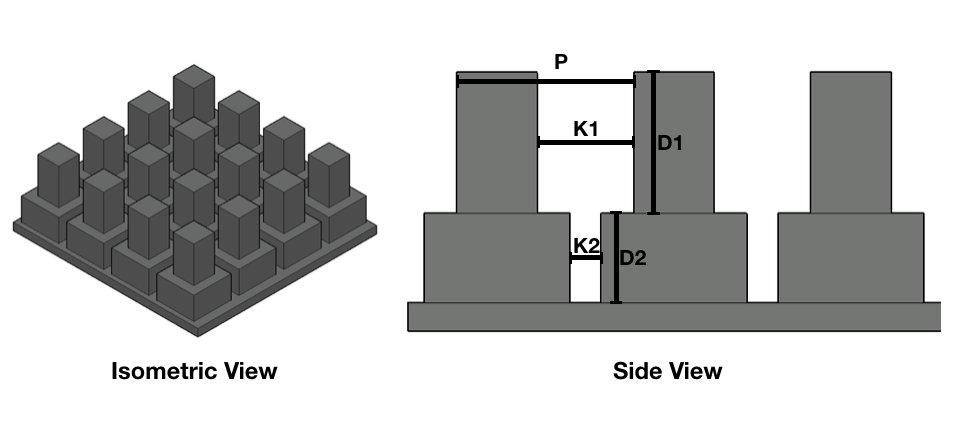}
\caption{(Left) Isometric view of a fiducial two-layer ARC design. (Right) Side view of the fiducial two-layer design with the relevant design parameters labeled.}
\label{fig:design}
\end{figure}

The design of the ARC is nearly identical in geometry to diced metamaterial ARCs that have been fabricated in silicon lenses and deployed on current CMB experiments \cite{Datta_Si_AR,Coughlin_si_AR,Golec_si_AR}. The ARC consists of nested cuts of fixed thickness and depth periodically diced into the optics surface as shown in Figure \ref{fig:design}. If the periodicity, or pitch (P), is sufficiently smaller than the wavelength of the incident radiation then the sub-wavelength metamaterial features do not scatter and behave like a traditional homogeneous layered AR coating \cite{Datta_Si_AR, pitch_criteria}.

The ARC design is modelled in the finite-element analysis software HFSS by Ansys \cite{Ansys_HFSS}. The initial design parameters are based on an existing two-layered metamaterial ARC design for silicon optics, but scaled to the alumina index and the 75-170 GHz MF band. The pitch is then fixed so that the coating will not scatter in the observing band at angles of incidence up to $20^\circ$ which is the most extreme angle that rays reflect off the alumina filters in the SO optical systems. The remaining design parameters, the widths (kerfs) and depths of the nested cuts, are then optimized to minimize reflection across the observing bands. 

The UHF design was numerically optimized for the 200-300 GHz band. In this case, the simple scaling of the ARC design was not possible due to blade limitations with regards to the thinner of the nested cuts. To overcome this issue, the kerf of the thinner cut (K2) was fixed to the smallest achievable kerf in alumina, 50 microns, and the optimization process adjusted the remaining three free parameters. In both the MF and UHF cases the optimal designs result in percent-level control of reflections across the bands which satisfies the design requirements for SO.

\begin{figure}[t!]
\centering
\includegraphics[width=\linewidth]{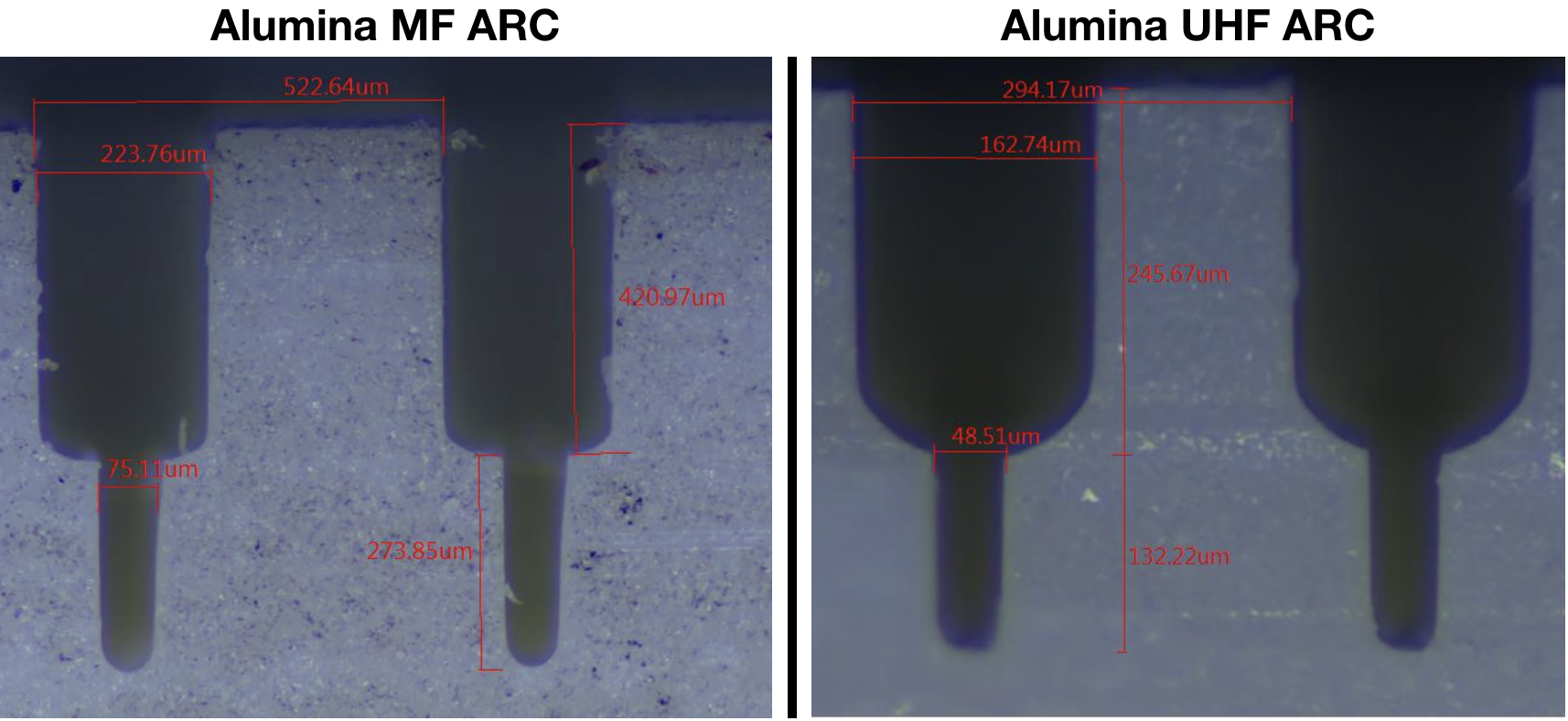}
\caption{(Left) Microscope image of the alumina MF ARC prototype. (Right) Microscope image of the alumina UHF ARC prototype.}
\label{fig:Microscope_Pictures}
\end{figure}

The initial optimal designs were prototyped and the measured profile of the cuts revealed that the dicing blade tool wear resulted in rounded rather than flat bottoms as shown in Figure \ref{fig:Microscope_Pictures}. This rounded geometry was incorporated into the HFSS simulation and the ARC designs re-optimized. The resulting final design parameters for the MF and UHF ARCs are summarized in table \ref{tab:DesignParams}.

\begin{table}[htbp]
\centering
\caption{Parameters of the two AR coating designs}
\begin{tabular}{ccc}
\hline
  & MF & UHF \\
\hline
Pitch (P)  & 0.522 mm  &  0.295 mm \\
Kerf 1 (K1)  & 0.220 mm  & 0.160 mm \\
Depth 1 (D1) & 0.425 mm  & 0.250 mm \\
Kerf 2 (K2)  & 0.070 mm  & 0.050 mm \\
Depth 2 (D2)  & 0.289 mm  & 0.138 mm \\
\hline
\end{tabular}
  \label{tab:DesignParams}
\end{table}

\begin{figure}[t!]
\centering
\includegraphics[width=\linewidth]{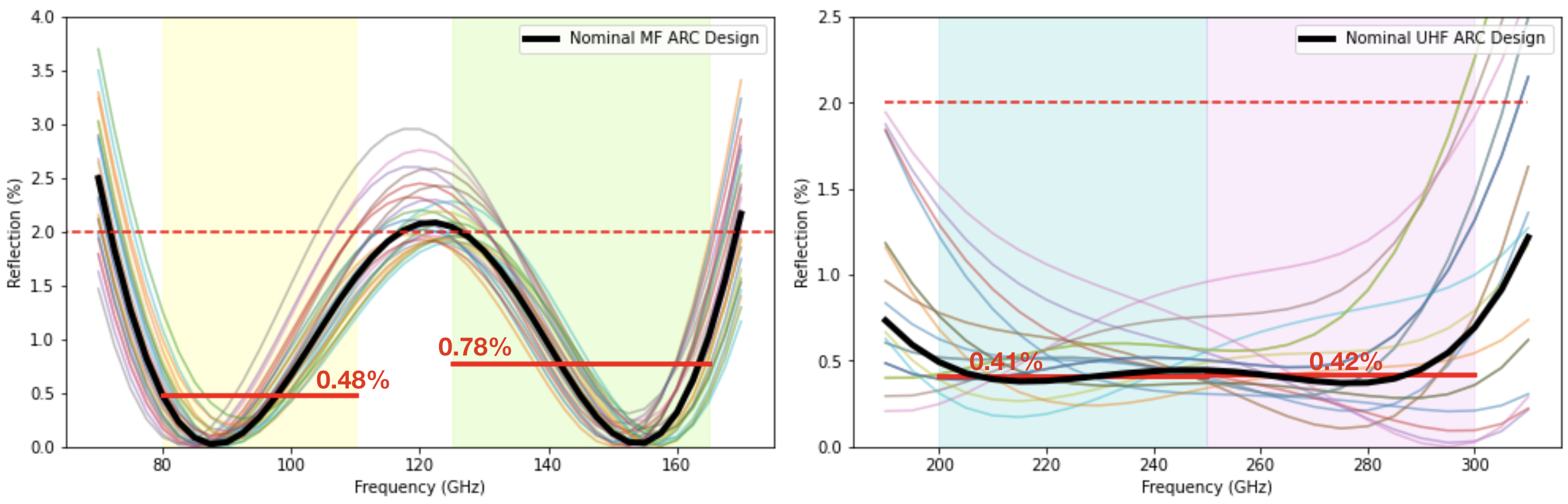}
\caption{(Left) Results of the MF ARC tolerancing analysis. (Right) Results of the UHF ARC tolerancing analysis. The semi-transparent curves denote variations from the nominal designed depths of the cuts. The solid red lines denote the average reflection across the bands and the red dotted lines denotes 2\% reflection.}
\label{fig:tolerance}
\end{figure}

After the optimal designs were chosen, the tolerance of the performance to errors in cut depth was analyzed. The dicing blade thicknesses are controlled to a three micron tolerance which has little to no effect on the ARC performance. Depth errors dominate the tolerance budget of the machining process. The ARCs were simulated at normal incidence with combinations of depth errors of $\pm20$ microns in steps of 10 microns for each effective layer. The results of this analysis, along with the nominal ARC performance, are shown in Figure \ref{fig:tolerance}. Note that the simulations were performed on a single sided ARC model and so the results represent the reflection per surface. All simulations that deviate from the nominal design are below 2\% reflection in the observing bands which is the design goal for SO and so we fix 20 microns as the requirement for depth control for the fabrication of the ARCs.  


\section{Fabrication}

\begin{figure}[t!]
\centering
\includegraphics[width=\linewidth]{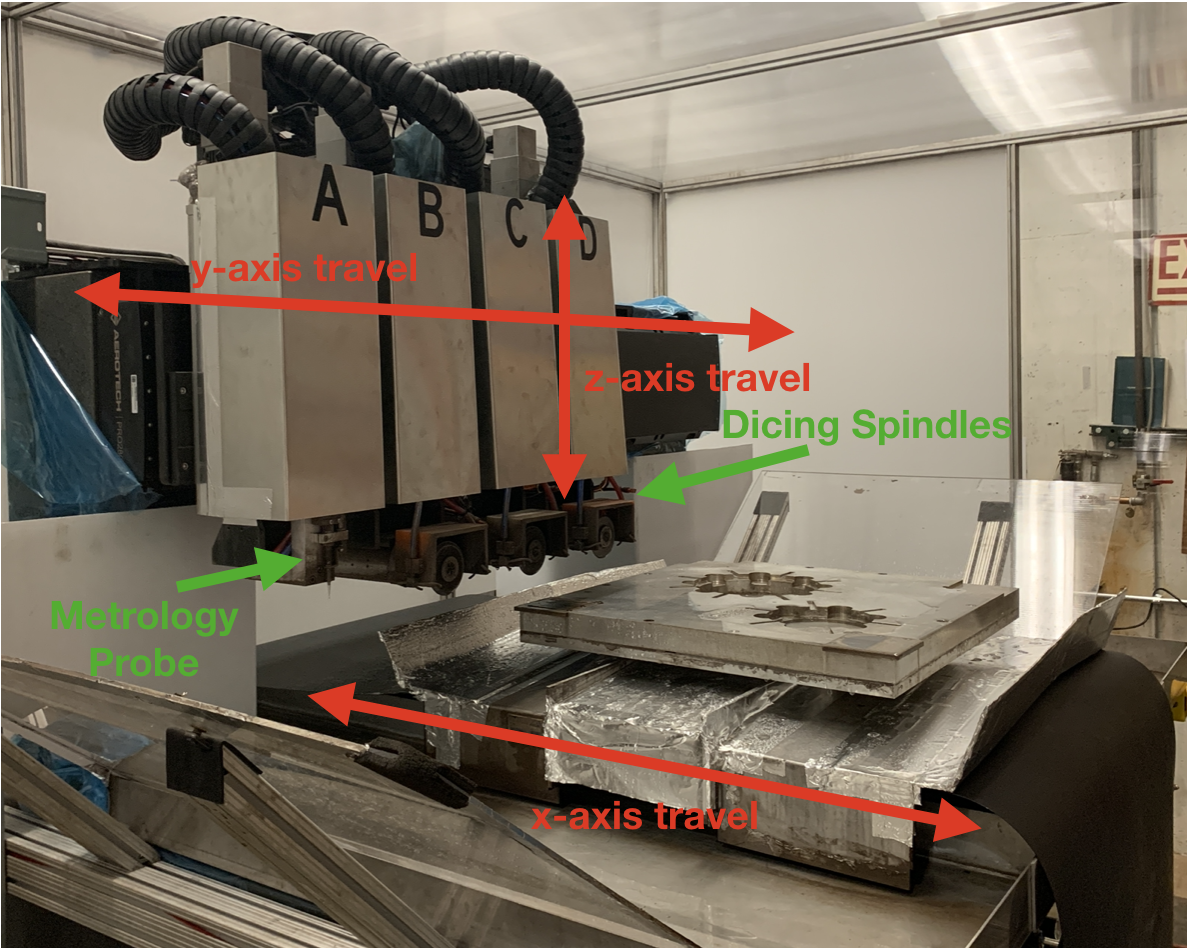}
\caption{An image of the dicing saw system used to fabricate metamaterial ARCs. The independent dicing spindles are labeled B, C, and D with the metrology axis labeled A.}
\label{fig:dicingsaw}
\end{figure}

The metamaterial ARCs were fabricated using a custom-built multi-axis dicing saw system built by our team for SO at Fermi National Accelerator Laboratory. This system (shown in Figure \ref{fig:dicingsaw}) consists of three dicing saw spindles mounted on independent  z-axes (100 mm of travel) on a common y-axis ($\pm 600$mm of travel). A ruby-tipped metrology probe with sub-micron accuracy is mounted on the forth z-axis. This probe is used to map the surface of the mounted optics to be cut. The filters to be machined are secured on a chuck mounted on a rotary stage with $360^\circ$ rotation on top of an x-axis ($\pm275$mm of travel). These axes allow the complete fabrication of an ARC on one side of an up to 600 mm diameter optic without need to dismount the optic.

The dicing procedure for the alumina ARC is similar to the process used for silicon \cite{Datta_Si_AR}, but because alumina is considerably harder than silicon extra steps must be taken to ensure the cuts maintain the correct depth. The procedure starts with mounting an alumina filter and using the metrology probe to map its surface. These data are fit to a flat plane model with Fourier corrections to account for deformations due to how the filter is clamped. The residuals of this fitting procedure are normally less than five microns. The cut trajectories are then generated for each dicing spindle. To confirm the absolute depth of the cuts, test touches are then performed with the dicing saw blades on the filter surface. These serve two purposes: the first is to confirm that the fit to the metrology data and the model is accurate; the second is to correct for any z-offset between the model surface and the actual filter which can arise from slight differences in the blade diameters.

A set of test touches are additionally performed on a test silicon wafer mounted to the same aluminum chuck as the alumina filter, which serves as a relative reference for blade wear.  After every 25 lines are diced on the filter, the system automatically makes a test touch on the silicon wafer. The test touches are chosen to be approximately 50 microns deep.  The length of the test touch represents a chord across the 78 mm diameter circle defined by our round blades.  Measurement of this chord allows a micron accurate determination of the depth of the cuts.  This measurement is used to determine the blade wear (diameter loss) as a function of millimeters of alumina diced.

Controlling for tool wear of the dicing blades is crucial for the success of these coatings. Different blades are used to realize the two different kerfs in both the MF and UHF ARCs. These blades differ not only in thickness, but also in how the zinc composite material they are made of is bonded and blade exposure which leads to multiple blade wear coefficients to track. Moreover, the blade wear also depends on which orientation of the ARC is being diced. Since the second set of cuts (rotated $90^\circ$ relative to the first set) dices through previously diced material, the blades need to remove roughly half the alumina and thus have a lower blade wear. For reference, for the MF ARC we have found that the most the blades wear on average is 6 microns per meter of alumina diced for the thick blade and 20 microns per meter for the thin blade. One 42 cm diameter MF filter requires roughly 260 meters of total length diced. With the maximum blade exposure without compromising cut quality, the dicing process requires roughly two thick and 12 thin blades to completely fabricate a filter. To streamline the fabrication process, we mount one of the thick blades on one spindle and thin blades on the remaining two spindles. This allows for longer cutting periods without blade changes. Every time the blade is changed the test touches on the silicon reference wafer calibrate the depth the new blades are cutting and allow new wear coefficients to be inferred.       

Using the procedures outlined in this section, four full-scale SO Large Aperture Telescope (LAT) MF filters were fabricated as well as a two-inch diameter UHF prototype. The ultimate production rate that was achieved for the MF filters (one of which is shown in Figure \ref{fig:MF_Filter}) was 15 days per 42 cm diameter filter. We believe that a higher rate can be achieved if the inspection of the test touches on the silicon reference wafer is automated. If a microscope is added to the system this test touch process could be done remotely or automated completely. This would allow for nearly continuous operation of the saw system and could cut the production rate to less than a week per filter.

\section{Metrology}

Throughout the development of the ARCs, measurements of the coating need to be made to confirm that the cuts are being diced to specification. We use several techniques to make these measurements including microscopy and photography.

Microscopy was performed on the smaller ARC prototypes since the larger filters do not fit in the available microscope. The microscope is also used during the fabrication process to inspect cuts diced into a test silicon wafer to ensure the dicing blades are not excessively worn such that the profile is compromised. Images taken with the microscope of the MF and UHF ARCs are shown in Figure \ref{fig:Microscope_Pictures} and one can clearly see that the profile is similar to the fiducial stepped pyramid design but with rounded bottoms due to the tool wear. The microscope images also confirm that the kerf of the cuts are to specification. The difference between the measured depths and the nominal values are within the 20 micron machining tolerance which confirms that the machining procedure outlined in the previous section is successful at achieving the design specifications for the ARCs.

\begin{figure}[t!]
\centering
\includegraphics[width=\linewidth]{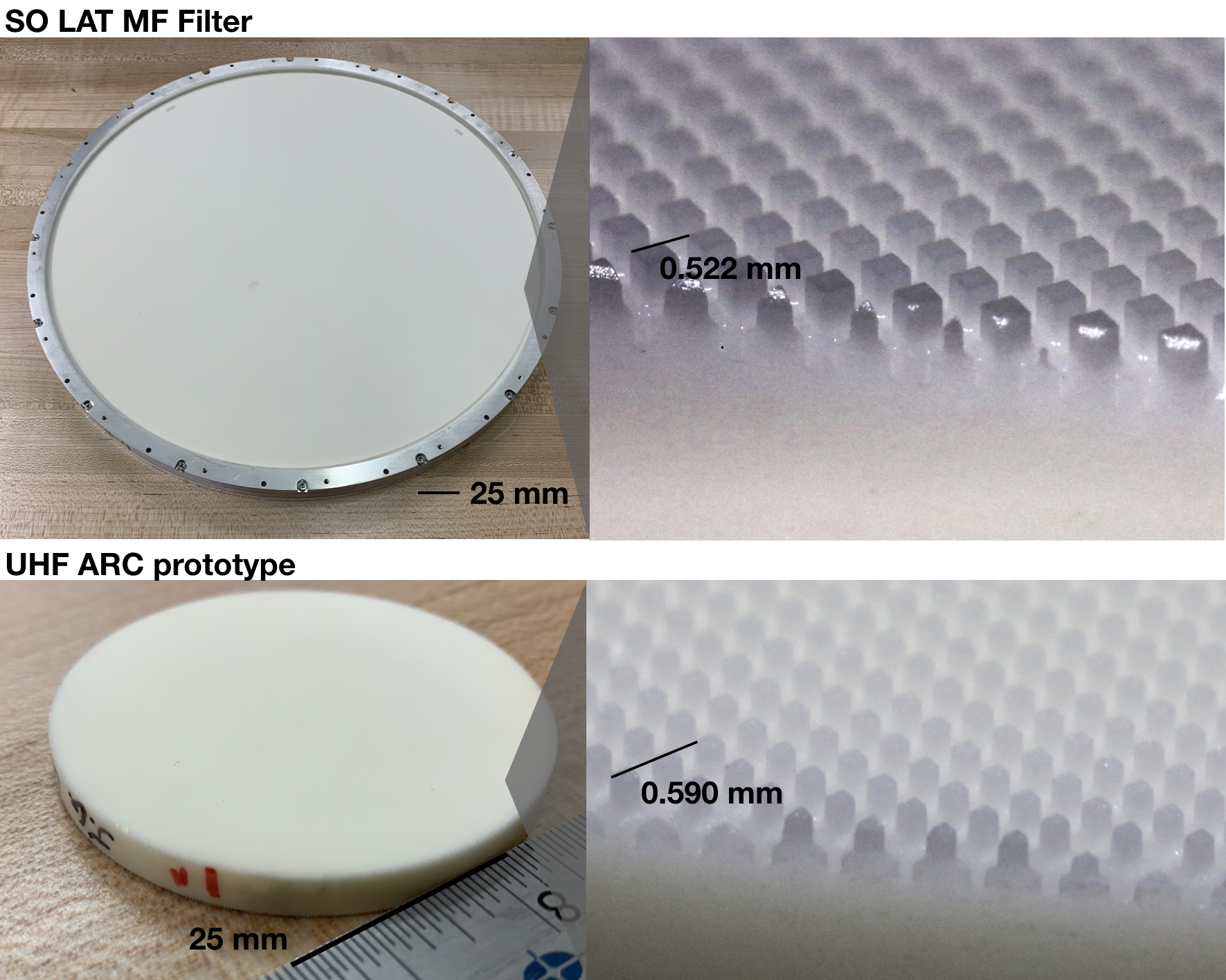}
\caption{(Top Left) Full sized image of an AR coated SO LAT MF filter. (Top Right) A zoomed in picture of the MF ARC. (Bottom Left) Full sized image of the UHF prototype ARC. (Bottom Right) A zoomed in picture of the UHF ARC.}
\label{fig:MF_Filter}
\end{figure}

We use a 5:1 macro lens on a digital camera to verify that there are no large scale variations in the ARCs and to inspect any defects that may appear. The resulting images from the camera serve as the quality assurance step in the fabrication procedure of the filters. The photos shown on the right side of Figure \ref{fig:MF_Filter} confirm that there is essentially no chipping or other defects that can be associated with the dicing procedure. Additionally, no defects appeared after thermally cycling the filter between room temperature and 80 Kelvin which is the operating temperature of these filters in the SO LAT cyrostat. Due to the robustness of alumina, the defect rate is extremely low with less than a dozen broken ARC pillars across the filter which contains over 500,000 pillars in total.

\section{Reflection and Transmission}

The reflection and transmission of the ARC filters and prototypes were measured using an ultra-broadband (20-1200 GHz) source and detector coupled to the sample using parabolic mirrors. The measurements presented here are made at ambient room temperature. The operating temperatures of the alumina filters in the SO optical systems are 80 K and below. While the index of refraction will not change appreciably from room temperature to those temperatures, the dielectric losses will decrease by as much as a factor of four \cite{Inoue_thermalandopticalproperties}. This will not to affect the reflection performance presented here, but it will result in improved in-band transmission.  The cryogenic dielectric loss was accounted for in the design of the IR blocking filters \cite{ningfeng_LATR_thermal}. 

\subsection{Transmitter and Receiver}
The ultra-broadband coherent source used to measure reflection and transmission uses two distributed feedback (DFB) lasers with indium gallium arsenide transmitter and receiver photomixers purchased from Toptica Photonics. The two DFB diode lasers send near-infrared frequencies to the emitter photomixer. The difference frequency from these two tuned lasers dictates the terahertz frequency generated in the photomixer. These two diode lasers are temperature controlled automatically by the accompanying digital controller. The emitter photomixer, i.e. the transmitter, outputs a beam with a divergence angle ranging from 12 to 15 degrees from a 25 mm diameter silicon lens encasing.

The transmitter (also referred to as the source) generates a continuous wave (cw) terahertz frequency signal. This photomixer is a metal-semiconductor-metal structure with interdigitated electrodes and a log-spiral antenna. The electrodes produce a photocurrent which oscillates at the difference frequency. This photocurrent is then output as our desired terahertz signal by the log-spiral antenna surrounding the photomixer.

The receiver photomixer, also encased in a silicon lens, is illuminated by both the terahertz wave and the laser beat. The photocurrent induced in the receiver photomixer is proportional to the amplitude of the signal's electric field \cite{terascan_manual}.

A fiber stretcher extension, which adjusts the optical path length difference between the transmitter and receiver, is used to modulate the phase at kHz frequencies.  The output of the receiver is then demodulated to get the amplitude and phase of the electric field. The minimum frequency resolution is 1 MHz when using the fiber stretcher and the source can sweep from 20-1200 GHz in 0.2 GHz steps in approximately thirteen minutes.

\begin{figure}
\centering
\includegraphics[width=2 in]{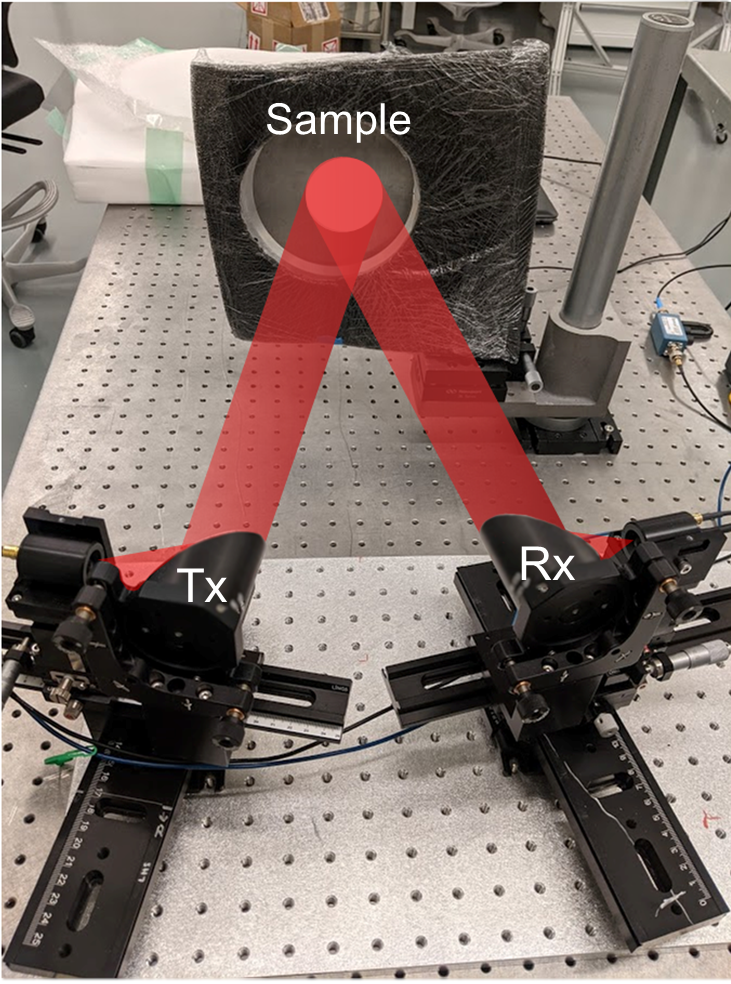}
\caption{\label{fig:refl_setup} The setup for the reflection measurement. The polished plate is pictured here in the sample holder. (Tx is the transmitter, Rx is the receiver.)} 
\end{figure}

\subsection{Reflection}
Figure \ref{fig:refl_setup} shows the setup for measuring reflection. Reflection measurements were made by mounting the sample on an adjustable mount that precisely aligns the sample to couple optimally between the source and receiver. The mount is made of aluminum and covered in an Eccosorb HR-10 layer to mitigate unwanted reflections. The setup of the system can support standing waves which reflect through the full optical path between the transmitter and receiver. These are reduced by two carbon-loaded polyethylene flat sheets, typically 0.5 mm thick (not pictured in Figure \ref{fig:refl_setup}), placed in the optical path just following the transmitter and receiver to work as attenuators \cite{black_poly}. These plates are oriented at $45^\circ$ relative to the beam propagation direction so that reflected light exits the system.

\begin{figure*}[h!]
\centering
\includegraphics[width=\textwidth]{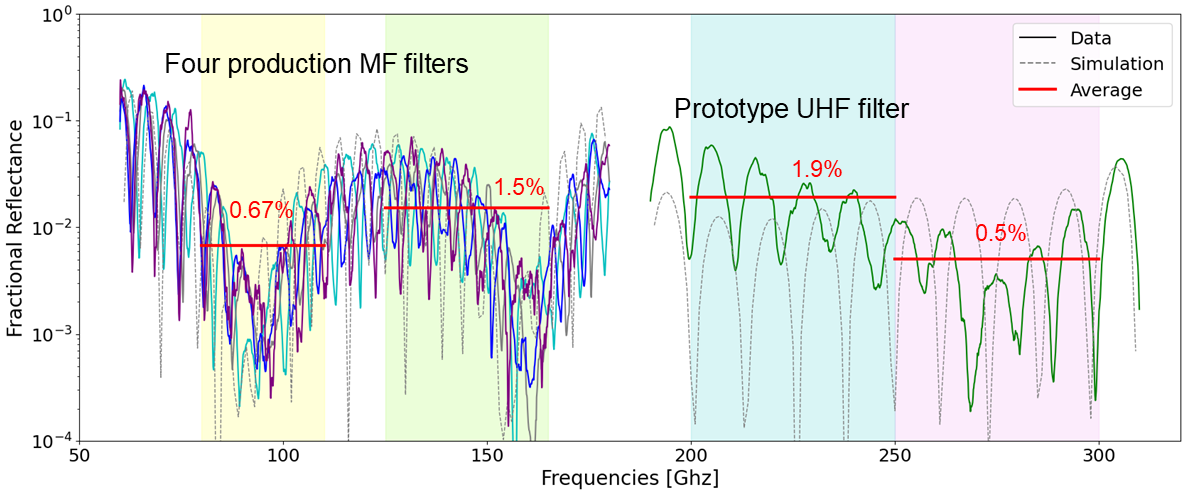}
\caption{\label{fig:allreflplot} Measured reflection of ARC of four 450 mm diameter MF filters and one 50 mm diameter prototype UHF filter. The red lines indicate the average reflection across the observing bands.} 
\end{figure*}

Calibration is done by measuring the reflected signal from a polished aluminum plate. The reflection from the sample is then measured with the surface of the ARC placed in the same optical plane as the aluminum plate. The ratio of the measurement and calibration are squared to find the fractional reflected power. While the phase measurements are stable at the degree level, these data are not considered here. The reflection was measured from 60-180 GHz for the MF alumina filters, and from 190-330 GHz for the UHF filter. The step size was set to 0.1 GHz with an integration time of 3.02 ms per frequency bin. The wedged MF filters are roughly 8 mm thick (the thickness varies between 5 and 11 mm) and the UHF filter prototype is 5 mm thick, so the 0.1 GHz frequency spacing of the reflection measurements is adequate to resolve fringing within the samples which should be on the order of several GHz. The measurements considered here were carried out at an angle of incidence of 14 degrees for the first MF filter and at 10 degrees for the other three MF and the UHF filters. 

The reflection measurements of four SO MF LAT filters and the prototype UHF filter are shown in Figure \ref{fig:allreflplot}. The four MF filters all have mean reflections less than 2\% in the observing bands which meets the SO design requirements for the alumina filters. Due to the wedge design of the filters, the reflection performance between all four filters are not identical since they were measured at different locations on the filter with varying thicknesses. However, all the filters produced are in line with HFSS simulations of the ARC. 

The UHF prototype also has a mean reflection less than 2\% across the UHF bands but performs slightly worse than predicted by the design HFSS simulation. Measurements of the ARC dimensions with a microscope revealed that the cuts were too deep by as much as 20 microns at points which caused poor performance in the lower frequency band. This error was confirmed post-fabrication by investigating the variability of the test touches made on the silicon reference wafer during ARC fabrication. Even with this error, the prototype achieves less than 2\% reflectance across this band. We anticipate improved UHF ARC performance when these filters enter production for SO since better active care will be taken to control the cut depths during fabrication. 


\subsection{Transmission} 
\begin{figure*}
 \centering
 \includegraphics[width=\textwidth]{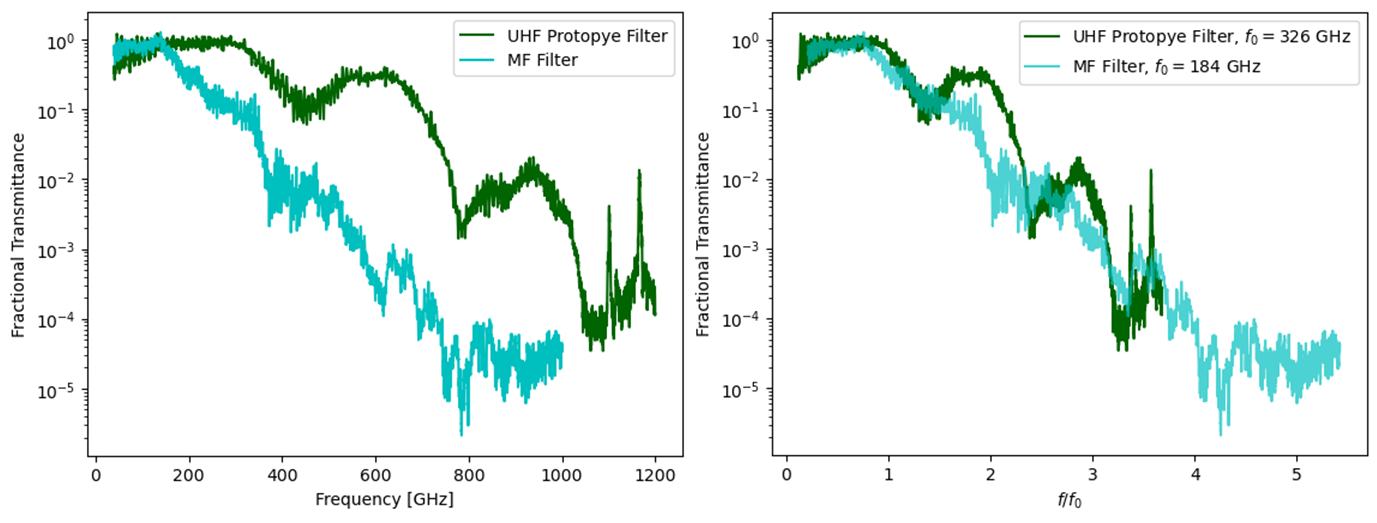}
 \caption{\label{fig:tmplot} Transmission of the UHF ARC alumina prototype and an MF ARC alumina production filter. (Left) The fractional fractional transmittance of the filters as a function of frequency. (Right) The transmittance as a function of scaled frequency where the frequencies are divided by $f_0$, the diffractive limit threshold frequency of the filter. This is done to highlight the transmission falling off at similar rates in both filters.} 
 \end{figure*}

The transmission was measured with the beam from the source aligned to point directly to the parabolic mirror pairing into the receiver. Calibration was performed by dividing the power received with the samples in place by the power received when the transmitter-to-receiver path is unblocked.  These measurements were carried out from 50 GHz to 1.2 THz in 0.1 GHz steps. The results are shown in Figure \ref{fig:tmplot} (see \cite{Yukithesis} for the transmission spectra of alumina without ARC).

The results for the MF and UHF coatings place a lower limit on in-band transmission at > 80\%.  The ability to assess transmission in-band are limited by the control of alignment and standing waves in the system which are not perfectly controlled. Given the reflectance measurements, and separate measurements of the loss-tangent of alumina \cite{Inoue_thermalandopticalproperties}, we interpreted this as consistent with 97\% transmission in-band.  

Both the MF and UHF coatings show a dramatic decrease in transmission above their respective bands. This apparent attenuation is more than an order of magnitude larger than what is expected from the loss tangent of the alumina material. This was confirmed by measuring the transmission of an uncoated alumina plate which showed no such attenuation at high frequency. We interpret this above band attenuation as the onset of scattering in these coatings which is anticipated as the wavelength of the incident light becomes comparable to the pitch of the ARC. In this wavelength regime, the ARC grating no longer satisfies the "quasi-static limit" condition where only the zeroth-order diffraction mode propagates and instead multi-mode propagation occurs \cite{Lalanne_EMT, Grann_EMT, Kikuta_EMT, pitch_criteria}. While a rigorous theoretical model that encompasses this behavior is beyond the scope of this paper, we can test our interpretation by scaling the frequency axis relative to the metamaterial breakdown frequency, or the frequency where the ARC grating enters the diffractive or multi-mode limit. This is given by \cite{pitch_criteria}
\begin{align}
    f_0 = \frac{c}{P(n_\text{alumina}+n_\text{vacuum} \sin \theta_i)}
\end{align}
where $P$ is the pitch of the ARC previously defined in section \hyperref[sec:Design]{2}, $n_\text{alumina}$ and $n_\text{vacuum}$ are the optical indices of alumina and vacuum respectively, $\theta_i$ is the angle of incidence, and $c$ is the speed of light. The diffractive limit threshold frequency as stated here is dependant on the pitch dimensions and therefore is a scale dependant property of the ARC. For the two filter designs presented in this work the diffractive limit threshold frequencies are $f_\text{0, MF} = 184 \text{ GHz}$ and $f_\text{0, UHF} = 326 \text{ GHz}$, for the MF and UHF filters respectively. We plot the transmission after scaling, $f_\text{scaled} = f/f_0$, in the right panel of Figure \ref{fig:tmplot}.  The qualitative agreement of these two measurements supports our interpretation with a change in transmission by approximately -10 dB per octave after entering the diffractive limit.
 
\section{Conclusion}

We have developed metamaterial ARCs for alumina filters for the SO MF and UHF bands. These ARCs achieve percent-level control of reflections across up to an octave of bandwidth, which is the best performance across the widest bandwidth for any alumina ARC technology to date. Four 420 mm diameter alumina filters for the SO LAT were fabricated with the MF ARC with an ultimate production rate of 15 days per filter. The reflection performance of these filters was measured and agrees with simulations. Additionally, a two-inch diameter prototype of the UHF ARC was fabricated and its reflection performance was measured to have percent-level reflections across the UHF band.

The transmission measurements indicate that these filters cause significant scattering above the target bands. If future work were to characterize the scattering kernel, this behavior could be exploited to help define the in-band filtering for future experiments. Since this scattering is out-of-band it is not  a liability in the current use cases. This was demonstrated with end-to-end measurements of a system including these filters which will be presented in a future publication \cite{Grace_holopaper,Sierra_LATRtpaper}.

In addition to the future work of the alumina UHF ARC on a full-scale filter, the MF ARC will be diced onto two alumina filters for the SO Small Aperture Telescope (SAT). The largest of these has a diameter of 550 mm which will be a further test for the alumina metamaterial ARC technology. We do not anticipate any production obstacles from the larger diameter except for a marginally longer production time. After those filters are finished, the remaining task for alumina ARC for SO concerns the remaining dichroic band, the 20/40 GHz (Low-Frequency or LF). An ARC design has been made for this final band, but details such as the dicing blade types and cutting procedure still remain to be determined.

The production rate and reliability of the metamaterial ARCs for alumina optics makes it a compelling technology for future millimeter and sub-millimeter experiments. Both laser ablated metamaterials fabricated for other experiments and the diced ARCs presented here have mechanical advantages that make them desirable for the cryogenic applications of alumina filters compared with the laminate AR coatings discussed in section \ref{sec:Introduction}. However the diced metamaterials have been demonstrated across a wider bandwidth, on larger-format optical filters, and at a much larger production scale than both laser machined ARCs and laminate coatings. These advantages show that diced metamaterial ARCs for alumina filters are a more mature technology ready for deployment on current generation experiments like SO, and deserve to be heavily considered as a baseline technology for future large-scale observatories like CMB-S4.  

While the work presented here is limited to flat surfaces, extending this technology to curved lenses is trivial since the same machine that produced the ARC for the alumina filters has also produced metamaterial ARCs for curved silicon lenses. Robust ARCs are critical to the present and future generation of millimeter wave experiments.   This work demonstrates that diced metamaterial ARCs provide a combination of good performance, extreme robustness, and a manageable production rate.

\begin{backmatter}
\bmsection{Funding} 

This work was funded by the Simons Foundation (Award \#457687, B.K.). JEG is supported by a NASA Space Technology Research Fellowship (80NSSC21K0411). SS is supported by a National Science Foundation Graduate Research Fellowship under Grant No. DGE 1746045.

\bmsection{Acknowledgments} 

This document was prepared by The Simons Observatory using the resources of the Fermi National Accelerator Laboratory (Fermilab), a U.S. Department of Energy, Office of Science, HEP User Facility. Fermilab is managed by Fermi Research Alliance, LLC (FRA), acting under Contract No. DE-AC02-07CH11359.

\bmsection{Disclosures} The authors declare no conflicts of interest.

\bmsection{Data availability}  Data underlying the results presented in this paper are not publicly available at this time but may be obtained from the authors upon reasonable request.

\end{backmatter}

\bibliography{sample}

\begin{thebibliography}{10}
\newcommand{\enquote}[1]{``#1''}

\bibitem{Ceratek_JP}
\url{https://www.ceratech.co.jp/en/product/material/alumina/a9951ld/}.

\bibitem{Alford_LowLoss}
N.~M. Alford and S.~J. Penn, \enquote{Sintered alumina with low dielectric
  loss,} {\protect\JournalTitle{Journal of Applied Physics}} \textbf{80},
  5895--5898 (1996).

\bibitem{Breeze_LowLoss}
J.~D. Breeze, X.~Aupi, and N.~M. Alford, \enquote{Ultralow loss polycrystalline
  alumina,} {\protect\JournalTitle{Applied Physics Letters}} \textbf{81},
  5021--5023 (2002).

\bibitem{Afsar_LowLoss}
M.~N. Afsar, \enquote{Precision millimeter-wave dielectric measurements of
  birefringent crystalline sapphire and ceramic alumina,}
  {\protect\JournalTitle{IEEE Transactions on Instrumentation and Measurement}}
  \textbf{IM-36}, 554--559 (1987).

\bibitem{Inoue_thermalandopticalproperties}
Y.~{Inoue}, N.~{Stebor}, P.~A.~R. {Ade}, Y.~{Akiba}, K.~{Arnold}, A.~E.
  {Anthony}, M.~{Atlas}, D.~{Barron}, A.~{Bender}, D.~{Boettger},
  J.~{Borrilll}, S.~{Chapman}, Y.~{Chinone}, A.~{Cukierman}, M.~{Dobbs},
  T.~{Elleflot}, J.~{Errard}, G.~{Fabbian}, C.~{Feng}, A.~{Gilbert}, N.~W.
  {Halverson}, M.~{Hasegawa}, K.~{Hattori}, M.~{Hazumi}, W.~L. {Holzapfel},
  Y.~{Hori}, G.~C. {Jaehnig}, A.~H. {Jaffe}, N.~{Katayama}, B.~{Keating},
  Z.~{Kermish}, R.~{Keskitalo}, T.~{Kisner}, M.~{Le Jeune}, A.~T. {Lee}, E.~M.
  {Leitch}, E.~{Linder}, F.~{Matsuda}, T.~{Matsumura}, X.~{Meng}, H.~{Morii},
  M.~J. {Myers}, M.~{Navaroli}, H.~{Nishino}, T.~{Okamura}, H.~{Paar},
  J.~{Peloton}, D.~{Poletti}, G.~{Rebeiz}, C.~L. {Reichardt}, P.~L. {Richards},
  C.~{Ross}, D.~E. {Schenck}, B.~D. {Sherwin}, P.~{Siritanasak}, G.~{Smecher},
  M.~{Sholl}, B.~{Steinbach}, R.~{Stompor}, A.~{Suzuki}, J.~{Suzuki},
  S.~{Takada}, S.~{Takakura}, T.~{Tomaru}, B.~{Wilson}, A.~{Yadav},
  H.~{Yamaguchi}, and O.~{Zahn}, \enquote{{Thermal and optical characterization
  for POLARBEAR-2 optical system},} in \emph{Millimeter, Submillimeter, and
  Far-Infrared Detectors and Instrumentation for Astronomy VII,}  vol. 9153 of
  \emph{Society of Photo-Optical Instrumentation Engineers (SPIE) Conference
  Series} W.~S. {Holland} and J.~{Zmuidzinas}, eds. (2014), p. 91533A.

\bibitem{Yukithesis}
Y.~Inoue, \enquote{{The thermal design of the POLARBEAR-2 experiment},}
  Master's thesis, Graduate University for Advanced Studies, Shonankokusaimura,
  Hayama, Miura District, Kanagawa 240-0193, Japan (2013).

\bibitem{Lamb_OpticalProperties}
J.~W. {Lamb}, \enquote{{Miscellaneous data on materials for millimetre and
  submillimetre optics},} {\protect\JournalTitle{International Journal of
  Infrared and Millimeter Waves}} \textbf{17}, 1997--2034 (1996).

\bibitem{AR_Coating_Review}
H.~K. Raut, V.~A. Ganesh, A.~S. Nair, and S.~Ramakrishna,
  \enquote{Anti-reflective coatings: A critical{,} in-depth review,}
  {\protect\JournalTitle{Energy Environ. Sci.}} \textbf{4}, 3779--3804 (2011).

\bibitem{SPT-3G_AR}
A.~Nadolski, J.~D. Vieira, J.~A. Sobrin, A.~M. Kofman, P.~A.~R. Ade, Z.~Ahmed,
  A.~J. Anderson, J.~S. Avva, R.~B. Thakur, A.~N. Bender, B.~A. Benson,
  L.~Bryant, J.~E. Carlstrom, F.~W. Carter, T.~W. Cecil, C.~L. Chang, J.~R.
  Cheshire, G.~E. Chesmore, J.~F. Cliche, A.~Cukierman, T.~de~Haan,
  M.~Dierickx, J.~Ding, D.~Dutcher, W.~Everett, J.~Farwick, K.~R. Ferguson,
  L.~Florez, A.~Foster, J.~Fu, J.~Gallicchio, A.~E. Gambrel, R.~W. Gardner,
  J.~C. Groh, S.~Guns, R.~Guyser, N.~W. Halverson, A.~H. Harke-Hosemann, N.~L.
  Harrington, R.~J. Harris, J.~W. Henning, W.~L. Holzapfel, D.~Howe, N.~Huang,
  K.~D. Irwin, O.~Jeong, M.~Jonas, A.~Jones, M.~Korman, J.~Kovac, D.~L. Kubik,
  S.~Kuhlmann, C.-L. Kuo, A.~T. Lee, A.~E. Lowitz, J.~McMahon, J.~Meier, S.~S.
  Meyer, D.~Michalik, J.~Montgomery, T.~Natoli, H.~Nguyen, G.~I. Noble,
  V.~Novosad, S.~Padin, Z.~Pan, P.~Paschos, J.~Pearson, C.~M. Posada, W.~Quan,
  A.~Rahlin, D.~Riebel, J.~E. Ruhl, J.~T. Sayre, E.~Shirokoff, G.~Smecher,
  A.~A. Stark, J.~Stephen, K.~T. Story, A.~Suzuki, C.~Tandoi, K.~L. Thompson,
  C.~Tucker, K.~Vanderlinde, G.~Wang, N.~Whitehorn, V.~Yefremenko, K.~W. Yoon,
  and M.~R. Young, \enquote{Broadband, millimeter-wave antireflection coatings
  for large-format, cryogenic aluminum oxide optics,}
  {\protect\JournalTitle{Appl. Opt.}} \textbf{59}, 3285--3295 (2020).

\bibitem{EpoxyAR}
Y.~Inoue, T.~Matsumura, M.~Hazumi, A.~T. Lee, T.~Okamura, A.~Suzuki, T.~Tomaru,
  and H.~Yamaguchi, \enquote{Cryogenic infrared filter made of alumina for use
  at millimeter wavelength,} {\protect\JournalTitle{Appl. Opt.}} \textbf{53},
  1727--1733 (2014).

\bibitem{Bicep_Plastic_Lam_AR}
M.~{Dierickx}, P.~A.~R. {Ade}, Z.~{Ahmed}, M.~{Amiri}, D.~{Barkats}, R.~{Basu
  Thakur}, C.~A. {Bischoff}, D.~{Beck}, J.~J. {Bock}, V.~{Buza}, J.~R.
  {Cheshire IV}, J.~{Connors}, J.~{Cornelison}, M.~{Crumrine}, A.~{Jozef
  Cukierman}, E.~{Denison}, L.~{Duband}, M.~{Eiben}, S.~{Fatigoni}, J.~P.
  {Filippini}, C.~{Giannakopoulos}, N.~{Goeckner-Wald}, D.~{Goldfinger}, J.~A.
  {Grayson}, P.~{Grimes}, G.~{Hall}, G.~{Halal}, M.~{Halpern}, E.~{Hand}, S.~A.
  {Harrison}, S.~{Henderson}, S.~{Hildebrandt}, G.~C. {Hilton}, J.~{Hubmayr},
  H.~{Hui}, K.~D. {Irwin}, J.~H. {Kang}, K.~S. {Karkare}, S.~{Kefeli}, J.~M.
  {Kovac}, C.-L. {Kuo}, K.~{Lau}, E.~M. {Leitch}, A.~{Lennox}, {K}, .~G.
  {Megerian}, L.~{Minutolo}, L.~{Moncelsi}, Y.~{Nakato}, T.~{Namikawa}, H.~T.
  {Nguyen}, R.~{O'brient}, S.~{Palladino}, M.~{Petroff}, N.~{Precup},
  T.~{Prouve}, C.~{Pryke}, B.~{Racine}, C.~D. {Reintsema}, D.~{Santalucia},
  A.~{Schillaci}, B.~{Schmitt}, B.~{Singari}, A.~{Soliman}, T.~{St Germaine},
  B.~{Steinbach}, R.~{Sudiwala}, K.~L. {Thompson}, C.~{Tucker}, A.~D. {Turner},
  C.~{Umilt{\`a}}, C.~{Verges}, A.~G. {Vieregg}, A.~{Wandui}, A.~C. {Weber},
  D.~{Wiebe}, J.~{Willmert}, W.~L.~K. {Wu}, H.-I. {Yang}, K.~W. {Yoon},
  E.~{Young}, C.~{Yu}, L.~{Zeng}, C.~{Zhang}, and S.~{Zhang}, \enquote{{Plastic
  Laminate Antireflective Coatings for Millimeter-wave Optics in BICEP Array},}
  {\protect\JournalTitle{arXiv e-prints}} arXiv:2111.14751 (2021).

\bibitem{Nitta_EpoxyARandLaserAR}
T.~{Nitta}, S.~{Sekiguchi}, Y.~{Sekimoto}, K.~{Mitsui}, N.~{Okada},
  K.~{Karatsu}, M.~{Naruse}, M.~{Sekine}, H.~{Matsuo}, T.~{Noguchi}, M.~{Seta},
  and N.~{Nakai}, \enquote{{Anti-reflection Coating for Cryogenic Silicon and
  Alumina Lenses in Millimeter-Wave Bands},} {\protect\JournalTitle{Journal of
  Low Temperature Physics}} \textbf{176}, 677--683 (2014).

\bibitem{Mullite_AR}
Y.~Inoue, T.~Hamada, M.~Hasegawa, M.~Hazumi, Y.~Hori, A.~Suzuki, T.~Tomaru,
  T.~Matsumura, T.~Sakata, T.~Minamoto, and T.~Hirai, \enquote{Two-layer
  anti-reflection coating with mullite and polyimide foam for large-diameter
  cryogenic infrared filters,} {\protect\JournalTitle{Appl. Opt.}} \textbf{55},
  D22--D28 (2016).

\bibitem{Laser_AR}
R.~{Takaku}, Q.~{Wen}, S.~{Cray}, M.~{Devlin}, S.~{Dicker}, S.~{Hanany},
  T.~{Hasebe}, T.~{Iida}, N.~{Katayama}, K.~{Konishi}, M.~{Kuwata-Gonokami},
  T.~{Matsumura}, N.~{Mio}, H.~{Sakurai}, Y.~{Sakurai}, R.~{Yamada}, and
  J.~{Yumoto}, \enquote{{Large diameter millimeter-wave low-pass filter made of
  alumina with laser ablated anti-reflection coating},}
  {\protect\JournalTitle{Optics Express}} \textbf{29}, 41745 (2021).

\bibitem{Matsumura_LaserAR}
T.~Matsumura, K.~Young, Q.~Wen, S.~Hanany, H.~Ishino, Y.~Inoue, M.~Hazumi,
  J.~Koch, O.~Suttman, and V.~Sch\"{u}tz, \enquote{Millimeter-wave broadband
  antireflection coatings using laser ablation of subwavelength structures,}
  {\protect\JournalTitle{Appl. Opt.}} \textbf{55}, 3502--3509 (2016).

\bibitem{Datta_Si_AR}
R.~Datta, C.~D. Munson, M.~D. Niemack, J.~J. McMahon, J.~Britton, E.~J.
  Wollack, J.~Beall, M.~J. Devlin, J.~Fowler, P.~Gallardo, J.~Hubmayr,
  K.~Irwin, L.~Newburgh, J.~P. Nibarger, L.~Page, M.~A. Quijada, B.~L. Schmitt,
  S.~T. Staggs, R.~Thornton, and L.~Zhang, \enquote{Large-aperture
  wide-bandwidth antireflection-coated silicon lenses for millimeter
  wavelengths,} {\protect\JournalTitle{Appl. Opt.}} \textbf{52}, 8747--8758
  (2013).

\bibitem{Coughlin_si_AR}
K.~P. Coughlin, J.~J. McMahon, K.~T. Crowley, B.~J. Koopman, K.~H. Miller,
  S.~M. Simon, and E.~J. Wollack, \enquote{Pushing the limits of broadband and
  high-frequency metamaterial silicon antireflection coatings,}
  {\protect\JournalTitle{Journal of Low Temperature Physics}} \textbf{193},
  876–885 (2018).

\bibitem{Golec_si_AR}
J.~E. {Golec}, J.~J. {McMahon}, A.~{Ali}, S.~{Dicker}, N.~{Galitzki},
  K.~{Harrington}, B.~{Westbrook}, E.~J. {Wollack}, Z.~{Xu}, and N.~{Zhu},
  \enquote{{Design and fabrication of metamaterial anti-reflection coatings for
  the Simons Observatory},} in \emph{Society of Photo-Optical Instrumentation
  Engineers (SPIE) Conference Series,}  vol. 11451 of \emph{Society of
  Photo-Optical Instrumentation Engineers (SPIE) Conference Series} (2020), p.
  114515T.

\bibitem{Thornton_ACT_Overview}
R.~J. Thornton, P.~A.~R. Ade, S.~Aiola, F.~E. Angil{\`{e}}, M.~Amiri, J.~A.
  Beall, D.~T. Becker, H.-M. Cho, S.~K. Choi, P.~Corlies, K.~P. Coughlin,
  R.~Datta, M.~J. Devlin, S.~R. Dicker, R.~Dünner, J.~W. Fowler, A.~E. Fox,
  P.~A. Gallardo, J.~Gao, E.~Grace, M.~Halpern, M.~Hasselfield, S.~W.
  Henderson, G.~C. Hilton, A.~D. Hincks, S.~P. Ho, J.~Hubmayr, K.~D. Irwin,
  J.~Klein, B.~Koopman, D.~Li, T.~Louis, M.~Lungu, L.~Maurin, J.~McMahon, C.~D.
  Munson, S.~Naess, F.~Nati, L.~Newburgh, J.~Nibarger, M.~D. Niemack,
  P.~Niraula, M.~R. Nolta, L.~A. Page, C.~G. Pappas, A.~Schillaci, B.~L.
  Schmitt, N.~Sehgal, J.~L. Sievers, S.~M. Simon, S.~T. Staggs, C.~Tucker,
  M.~Uehara, J.~van Lanen, J.~T. Ward, and E.~J. Wollack, \enquote{{THE}
  {ATACAMA} {COSMOLOGY} {TELESCOPE}: {THE} {POLARIZATION}-{SENSITIVE} {ACTPol}
  {INSTRUMENT},} {\protect\JournalTitle{The Astrophysical Journal Supplement
  Series}} \textbf{227}, 21 (2016).

\bibitem{CLASS_Design_Overview}
J.~{Iuliano}, J.~{Eimer}, L.~{Parker}, G.~{Rhoades}, A.~{Ali}, J.~W. {Appel},
  C.~{Bennett}, M.~{Brewer}, R.~{Bustos}, D.~{Chuss}, J.~{Cleary}, J.~{Couto},
  S.~{Dahal}, K.~{Denis}, R.~{D{\"u}nner}, T.~{Essinger-Hileman}, P.~{Fluxa},
  M.~{Halpern}, K.~{Harrington}, K.~{Helson}, G.~{Hilton}, G.~{Hinshaw},
  J.~{Hubmayr}, J.~{Karakla}, T.~{Marriage}, N.~{Miller}, J.~J. {McMahon},
  C.~{Nu{\~n}ez}, I.~{Padilla}, G.~{Palma}, M.~{Petroff}, B.~{Pradenas
  M{\'a}rquez}, R.~{Reeves}, C.~{Reintsema}, K.~{Rostem}, D.~{Augusto Nunes
  Valle}, T.~{Van Engelhoven}, B.~{Wang}, Q.~{Wang}, D.~{Watts}, J.~{Weiland},
  E.~J. {Wollack}, Z.~{Xu}, Z.~{Yan}, and L.~{Zeng}, \enquote{{The Cosmology
  Large Angular Scale Surveyor receiver design},} in \emph{Millimeter,
  Submillimeter, and Far-Infrared Detectors and Instrumentation for Astronomy
  IX,}  vol. 10708 of \emph{Society of Photo-Optical Instrumentation Engineers
  (SPIE) Conference Series} J.~{Zmuidzinas} and J.-R. {Gao}, eds. (2018), p.
  1070828.

\bibitem{TolTEC_Overview}
G.~W. {Wilson}, S.~{Abi-Saad}, P.~{Ade}, I.~{Aretxaga}, J.~{Austermann},
  Y.~{Ban}, J.~{Bardin}, J.~{Beall}, M.~{Berthoud}, S.~{Bryan}, J.~{Bussan},
  E.~{Castillo}, M.~{Chavez}, R.~{Contente}, N.~S. {DeNigris}, B.~{Dober},
  M.~{Eiben}, D.~{Ferrusca}, L.~{Fissel}, J.~{Gao}, J.~E. {Golec}, R.~{Golina},
  A.~{Gomez}, S.~{Gordon}, R.~{Gutermuth}, G.~{Hilton}, M.~{Hosseini},
  J.~{Hubmayr}, D.~{Hughes}, S.~{Kuczarski}, D.~{Lee}, E.~{Lunde}, Z.~{Ma},
  H.~{Mani}, P.~{Mauskopf}, M.~{McCrackan}, C.~{McKenney}, J.~{McMahon},
  G.~{Novak}, G.~{Pisano}, A.~{Pope}, A.~{Ralston}, I.~{Rodriguez},
  D.~{S{\'a}nchez-Arg{\"u}elles}, F.~P. {Schloerb}, S.~{Simon}, A.~{Sinclair},
  K.~{Souccar}, A.~{Torres Campos}, C.~{Tucker}, J.~{Ullom}, E.~{Van Camp},
  J.~{Van Lanen}, M.~{Velazquez}, M.~{Vissers}, E.~{Weeks}, and M.~S. {Yun},
  \enquote{{The TolTEC camera: an overview of the instrument and in-lab testing
  results},} in \emph{Society of Photo-Optical Instrumentation Engineers (SPIE)
  Conference Series,}  vol. 11453 of \emph{Society of Photo-Optical
  Instrumentation Engineers (SPIE) Conference Series} (2020), p. 1145302.

\bibitem{SO_Instrument_Overview}
N.~{Galitzki}, A.~{Ali}, K.~S. {Arnold}, P.~C. {Ashton}, J.~E. {Austermann},
  C.~{Baccigalupi}, T.~{Baildon}, D.~{Barron}, J.~A. {Beall}, S.~{Beckman},
  S.~M.~M. {Bruno}, S.~{Bryan}, P.~G. {Calisse}, G.~E. {Chesmore},
  Y.~{Chinone}, S.~K. {Choi}, G.~{Coppi}, K.~D. {Crowley}, K.~T. {Crowley},
  A.~{Cukierman}, M.~J. {Devlin}, S.~{Dicker}, B.~{Dober}, S.~M. {Duff},
  J.~{Dunkley}, G.~{Fabbian}, P.~A. {Gallardo}, M.~{Gerbino},
  N.~{Goeckner-Wald}, J.~E. {Golec}, J.~E. {Gudmundsson}, E.~E. {Healy},
  S.~{Henderson}, C.~A. {Hill}, G.~C. {Hilton}, S.-P.~P. {Ho}, L.~A. {Howe},
  J.~{Hubmayr}, O.~{Jeong}, B.~{Keating}, B.~J. {Koopman}, K.~{Kiuchi},
  A.~{Kusaka}, J.~{Lashner}, A.~T. {Lee}, Y.~{Li}, M.~{Limon}, M.~{Lungu},
  F.~{Matsuda}, P.~D. {Mauskopf}, A.~J. {May}, N.~{McCallum}, J.~{McMahon},
  F.~{Nati}, M.~D. {Niemack}, J.~L. {Orlowski-Scherer}, S.~C. {Parshley},
  L.~{Piccirillo}, M.~{Sathyanarayana Rao}, C.~{Raum}, M.~{Salatino}, J.~S.
  {Seibert}, C.~{Sierra}, M.~{Silva-Feaver}, S.~M. {Simon}, S.~T. {Staggs},
  J.~R. {Stevens}, A.~{Suzuki}, G.~{Teply}, R.~{Thornton}, C.~{Tsai}, J.~N.
  {Ullom}, E.~M. {Vavagiakis}, M.~R. {Vissers}, B.~{Westbrook}, E.~J.
  {Wollack}, Z.~{Xu}, and N.~{Zhu}, \enquote{{The Simons Observatory:
  instrument overview},} in \emph{Millimeter, Submillimeter, and Far-Infrared
  Detectors and Instrumentation for Astronomy IX,}  vol. 10708 of \emph{Society
  of Photo-Optical Instrumentation Engineers (SPIE) Conference Series}
  J.~{Zmuidzinas} and J.-R. {Gao}, eds. (2018), p. 1070804.

\bibitem{CMBS4_sciencebook}
K.~N. {Abazajian}, P.~{Adshead}, Z.~{Ahmed}, S.~W. {Allen}, D.~{Alonso}, K.~S.
  {Arnold}, C.~{Baccigalupi}, J.~G. {Bartlett}, N.~{Battaglia}, B.~A. {Benson},
  C.~A. {Bischoff}, J.~{Borrill}, V.~{Buza}, E.~{Calabrese}, R.~{Caldwell},
  J.~E. {Carlstrom}, C.~L. {Chang}, T.~M. {Crawford}, F.-Y. {Cyr-Racine},
  F.~{De Bernardis}, T.~{de Haan}, S.~{di Serego Alighieri}, J.~{Dunkley},
  C.~{Dvorkin}, J.~{Errard}, G.~{Fabbian}, S.~{Feeney}, S.~{Ferraro}, J.~P.
  {Filippini}, R.~{Flauger}, G.~M. {Fuller}, V.~{Gluscevic}, D.~{Green},
  D.~{Grin}, E.~{Grohs}, J.~W. {Henning}, J.~C. {Hill}, R.~{Hlozek},
  G.~{Holder}, W.~{Holzapfel}, W.~{Hu}, K.~M. {Huffenberger}, R.~{Keskitalo},
  L.~{Knox}, A.~{Kosowsky}, J.~{Kovac}, E.~D. {Kovetz}, C.-L. {Kuo},
  A.~{Kusaka}, M.~{Le Jeune}, A.~T. {Lee}, M.~{Lilley}, M.~{Loverde}, M.~S.
  {Madhavacheril}, A.~{Mantz}, D.~J.~E. {Marsh}, J.~{McMahon}, P.~D.
  {Meerburg}, J.~{Meyers}, A.~D. {Miller}, J.~B. {Munoz}, H.~N. {Nguyen}, M.~D.
  {Niemack}, M.~{Peloso}, J.~{Peloton}, L.~{Pogosian}, C.~{Pryke}, M.~{Raveri},
  C.~L. {Reichardt}, G.~{Rocha}, A.~{Rotti}, E.~{Schaan}, M.~M. {Schmittfull},
  D.~{Scott}, N.~{Sehgal}, S.~{Shandera}, B.~D. {Sherwin}, T.~L. {Smith},
  L.~{Sorbo}, G.~D. {Starkman}, K.~T. {Story}, A.~{van Engelen}, J.~D.
  {Vieira}, S.~{Watson}, N.~{Whitehorn}, and W.~L. {Kimmy Wu}, \enquote{{CMB-S4
  Science Book, First Edition},} {\protect\JournalTitle{arXiv e-prints}}
  arXiv:1610.02743 (2016).

\bibitem{CMBS4_techbook}
M.~H. {Abitbol}, Z.~{Ahmed}, D.~{Barron}, R.~{Basu Thakur}, A.~N. {Bender},
  B.~A. {Benson}, C.~A. {Bischoff}, S.~A. {Bryan}, J.~E. {Carlstrom}, C.~L.
  {Chang}, D.~T. {Chuss}, K.~T. {Crowley}, A.~{Cukierman}, T.~{de Haan},
  M.~{Dobbs}, T.~{Essinger-Hileman}, J.~P. {Filippini}, K.~{Ganga}, J.~E.
  {Gudmundsson}, N.~W. {Halverson}, S.~{Hanany}, S.~W. {Henderson}, C.~A.
  {Hill}, S.-P.~P. {Ho}, J.~{Hubmayr}, K.~{Irwin}, O.~{Jeong}, B.~R. {Johnson},
  S.~A. {Kernasovskiy}, J.~M. {Kovac}, A.~{Kusaka}, A.~T. {Lee}, S.~{Maria},
  P.~{Mauskopf}, J.~J. {McMahon}, L.~{Moncelsi}, A.~W. {Nadolski}, J.~M.
  {Nagy}, M.~D. {Niemack}, R.~C. {O'Brient}, S.~{Padin}, S.~C. {Parshley},
  C.~{Pryke}, N.~A. {Roe}, K.~{Rostem}, J.~{Ruhl}, S.~M. {Simon}, S.~T.
  {Staggs}, A.~{Suzuki}, E.~R. {Switzer}, O.~{Tajima}, K.~L. {Thompson},
  P.~{Timbie}, G.~S. {Tucker}, J.~D. {Vieira}, A.~G. {Vieregg}, B.~{Westbrook},
  E.~J. {Wollack}, K.~W. {Yoon}, K.~S. {Young}, and E.~Y. {Young},
  \enquote{{CMB-S4 Technology Book, First Edition},}
  {\protect\JournalTitle{arXiv e-prints}} arXiv:1706.02464 (2017).

\bibitem{pitch_criteria}
D.~H. Raguin and G.~M. Morris, \enquote{Analysis of antireflection-structured
  surfaces with continuous one-dimensional surface profiles,}
  {\protect\JournalTitle{Appl. Opt.}} \textbf{32}, 2582--2598 (1993).

\bibitem{Ansys_HFSS}
\url{https://www.ansys.com/products/electronics/ansys-hfss}.

\bibitem{ningfeng_LATR_thermal}
N.~{Zhu}, T.~{Bhandarkar}, G.~{Coppi}, A.~M. {Kofman}, J.~L.
  {Orlowski-Scherer}, Z.~{Xu}, S.~{Adachi}, P.~{Ade}, S.~{Aiola},
  J.~{Austermann}, A.~O. {Bazarko}, J.~A. {Beall}, S.~{Bhimani}, J.~R. {Bond},
  G.~E. {Chesmore}, S.~K. {Choi}, J.~{Connors}, N.~F. {Cothard}, M.~{Devlin},
  S.~{Dicker}, B.~{Dober}, C.~J. {Duell}, S.~M. {Duff}, R.~{D{\"u}nner},
  G.~{Fabbian}, N.~{Galitzki}, P.~A. {Gallardo}, J.~E. {Golec}, S.~K.
  {Haridas}, K.~{Harrington}, E.~{Healy}, S.-P.~P. {Ho}, Z.~B. {Huber},
  J.~{Hubmayr}, J.~{Iuliano}, B.~R. {Johnson}, B.~{Keating}, K.~{Kiuchi}, B.~J.
  {Koopman}, J.~{Lashner}, A.~T. {Lee}, Y.~{Li}, M.~{Limon}, M.~{Link}, T.~J.
  {Lucas}, H.~{McCarrick}, J.~{Moore}, F.~{Nati}, L.~B. {Newburgh}, M.~D.
  {Niemack}, E.~{Pierpaoli}, M.~J. {Randall}, K.~P. {Sarmiento}, L.~J.
  {Saunders}, J.~{Seibert}, C.~{Sierra}, R.~{Sonka}, J.~{Spisak},
  S.~{Sutariya}, O.~{Tajima}, G.~P. {Teply}, R.~J. {Thornton}, T.~{Tsan},
  C.~{Tucker}, J.~{Ullom}, E.~M. {Vavagiakis}, M.~R. {Vissers}, S.~{Walker},
  B.~{Westbrook}, E.~J. {Wollack}, and M.~{Zannoni}, \enquote{{The Simons
  Observatory Large Aperture Telescope Receiver},} {\protect\JournalTitle{The
  Astrophysical Journal Supplement Series}} \textbf{256}, 23 (2021).

\bibitem{terascan_manual}
Toptica Photonics AG, Lochhamer Schlag 19, D-82166 Graefelfing/Munich,
  \emph{TeraScan 1550/TeraBeam 1550 Manual}, 03rd ed. (2019).

\bibitem{black_poly}
\enquote{{Carbon-Filled Slippery UHMW Polyethylene Film},}
  \url{https://www.mcmaster.com/8327K13/}. Accessed: 2022-06-27.

\bibitem{Lalanne_EMT}
P.~Lalanne and D.~Lemercier-lalanne, \enquote{On the effective medium theory of
  subwavelength periodic structures,} {\protect\JournalTitle{Journal of Modern
  Optics}} \textbf{43}, 2063--2085 (1996).

\bibitem{Grann_EMT}
E.~B. Grann, M.~G. Moharam, and D.~A. Pommet, \enquote{Artificial uniaxial and
  biaxial dielectrics with use of two-dimensional subwavelength binary
  gratings,} {\protect\JournalTitle{J. Opt. Soc. Am. A}} \textbf{11},
  2695--2703 (1994).

\bibitem{Kikuta_EMT}
H.~Kikuta, Y.~Ohira, H.~Kubo, and K.~Iwata, \enquote{Effective medium theory of
  two-dimensional subwavelength gratings in the non-quasi-static limit,}
  {\protect\JournalTitle{J. Opt. Soc. Am. A}} \textbf{15}, 1577--1585 (1998).

\bibitem{Grace_holopaper}
G.~Chesmore. Department of Physics, University of Chicago, 5720 S Ellis Ave,
  Chicago, IL 60637 is preparing a manuscript to be called "The Simons
  Observatory: Characterizing the Large Aperture Telescope Receiver with Radio
  Holography".

\bibitem{Sierra_LATRtpaper}
C.~Sierra. Department of Physics, University of Chicago, 5720 S Ellis Ave,
  Chicago, IL 60637 is preparing a manuscript to be called "Simons Observatory:
  Optical Characterization and Predicted Performance of the Mid-Frequency Large
  Aperture Telescope Optics".

\end{thebibliography}


\ifthenelse{\equal{\journalref}{aop}}{%
\section*{Author Biographies}
\begingroup
\setlength\intextsep{0pt}
\begin{minipage}[t][6.3cm][t]{1.0\textwidth} 
  \begin{wrapfigure}{L}{0.25\textwidth}
    \includegraphics[width=0.25\textwidth]{john_smith.eps}
  \end{wrapfigure}
  \noindent
  {\bfseries John Smith} received his BSc (Mathematics) in 2000 from The University of Maryland. His research interests include lasers and optics.
\end{minipage}
\begin{minipage}{1.0\textwidth}
  \begin{wrapfigure}{L}{0.25\textwidth}
    \includegraphics[width=0.25\textwidth]{alice_smith.eps}
  \end{wrapfigure}
  \noindent
  {\bfseries Alice Smith} also received her BSc (Mathematics) in 2000 from The University of Maryland. Her research interests also include lasers and optics.
\end{minipage}
\endgroup
}{}

\end{document}